\newif{\ifdraft}\drafttrue
\draftfalse
\ifdraft
\documentclass[draft,oribibl]{llncs}
\usepackage[notcite,color]{showkeys}
\else
\documentclass[oribibl,final]{llncs}
\fi 
\usepackage{etex}
\usepackage[applemac]{inputenc}
\usepackage{amsmath}
\usepackage{amssymb}
\usepackage{theorem}
\usepackage{lmodern}
\usepackage{fontenc}[T1]
\usepackage{enumerate}
\usepackage{enumitem}
\usepackage[english]{babel}
\usepackage{ifthen}

\usepackage{multirow}
\usepackage{xspace}
\usepackage{verbatim}

\usepackage[ruled]{algorithm}
\usepackage[noend]{algpseudocode}

\usepackage[final]{hyperref}

\usepackage{pdfsync}

\DeclareMathOperator{\End}{End}

\usepackage[ruled]{algorithm}
\usepackage[noend]{algpseudocode}

\usepackage{tikz}
\usetikzlibrary{arrows,decorations.markings}

\definecolor{darkgreen}{rgb}{0.0,0.7,0.0}
\newenvironment{LC}{\noindent\color{darkgreen} LC:}{}

\newenvironment{vd}{\noindent\color{blue} \colorbox{blue}{\color{black} VD:}}{}

\newenvironment{ME}{\noindent\color{magenta} ME:} {}

\usepackage{prettyref}
\newcommand{\prref}[1]{\prettyref{#1}}
\newrefformat{thm}{Theorem~\ref{#1}}
\newrefformat{lem}{Lemma~\ref{#1}}
\newrefformat{def}{Definition~\ref{#1}}
\newrefformat{cor}{Corollary~\ref{#1}}
\newrefformat{prop}{Proposition~\ref{#1}}
\newrefformat{sec}{Section~\ref{#1}}
\newrefformat{subsec}{Subsection~\ref{#1}}
\newrefformat{kap}{Chapter~\ref{#1}}
\newrefformat{ex}{Example~\ref{#1}}
\newrefformat{rem}{Remark~\ref{#1}}
\newrefformat{fig}{Figure~\ref{#1}}
\newrefformat{app}{Appendix~\ref{#1}}
\newrefformat{eq}{Equation~(\ref{#1})}
\newrefformat{tab}{Table~\ref{#1}}
\newrefformat{ap}{{\sc Appendix}}

\theoremstyle{plain}

\newcommand{\Dio}{Diophantine\xspace}

\newcommand{\solu}{solution\xspace}

\newcommand{\IFF}{if and only if\xspace}
\renewcommand{\hom}{homomorphism\xspace}

\newcommand{\Endo}{endomorphism\xspace}
\newcommand{\Endos}{endomorphisms\xspace}

\usepackage[framemethod=tikz]{mdframed} 

\newcommand{\set}[2]{\left\{#1\mathrel{\left|\vphantom{#1}\vphantom{#2}\right.}#2\right\}}

\newcommand{\oneset}[1]{\left\{\mathinner{#1}\right\}}
\newcommand{\os}{\oneset}

\newcommand{\es}{\emptyset}
\newcommand{\sse}{\subseteq}

\newcommand{\vdmatrix}[4]{\left(\begin{smallmatrix}#1 & #2\\ #3 & #4\end{smallmatrix}\right)}

\newcommand{\abs}[1]{\left|\mathinner{#1}\right|}

\newcommand{\Absone}[1]{\left\Vert\mathinner{#1}\right\Vert_{1}}

\newcommand{\N}{\ensuremath{\mathbb{N}}}
\newcommand{\Z}{\ensuremath{\mathbb{Z}}}

\newcommand{\PSPACE}{\ensuremath{\mathsf{PSPACE}}}

\newcommand{\NP}{\ensuremath{\mathsf{NP}}}
\newcommand{\NSPACE}{\ensuremath{\mathsf{NSPACE}}}
\newcommand{\EXPSPACE}{\ensuremath{\mathsf{EXPSPACE}}}

\renewcommand{\phi}{\varphi}

\newcommand{\lam}{\lambda}
\newcommand{\sig}{\sigma}

\newcommand\OO{\Omega}

\newcommand\SL{\mathop\mathrm{SL}}
\newcommand\PSL{\mathop\mathrm{PSL}}

\newcommand{\cA}{\mathcal{A}}

\newcommand{\cR}{\mathcal{R}}

\newif{\ifsecappendix}\secappendixtrue
\newif{\ifsecquestion}\secquestiontrue
\newif{\ifsecoldstuff}\secoldstufftrue
\newif{\ifAlles}\Allestrue

\Allesfalse 
\ifAlles\else
\secappendixfalse 
\secquestionfalse  
\secoldstufffalse  
\fi

\begin{document}

\pagestyle{plain}

\title{More Than 1700 Years of Word Equations}
\author{Volker Diekert}
\authorrunning{V.~Diekert}

\institute{Institut f\"ur Formale Methoden der Informatik,
  Universit\"at Stuttgart, Germany}

\maketitle

\begin{abstract}
Geometry and Diophantine equations have been ever-present in mathematics.  Diophantus of Alexandria was born in the 3rd century (as far as we know), but a systematic mathematical study of word equations began only in the 20th century. So, the title of the present article
does not seem to be justified at all. However, a linear Diophantine equation can be viewed as a special case of a system of word equations over a unary alphabet, and, more importantly, 
a word equation can be viewed as a special case of a Diophantine equation. Hence,
the problem WordEquations: ``Is a given word equation solvable?'', is intimately 
 related to Hilbert's 10th problem 
 on the solvability of Diophantine equations.  
 This became clear to the Russian school of mathematics at the latest in the mid 1960s, after which a systematic study of that relation began. 

Here, we review some recent developments which led to an amazingly simple decision procedure for WordEquations, and to the description of the set of all solutions 
as an EDT0L language. 
 \end{abstract}

\section*{Word Equations}
A word equation is easy to describe: it is a pair $(U,V)$ where 
$U$ and $V$ are strings over finite sets of constants $A$ and variables 
$\OO$. A \emph{solution} is mapping 
$\sig: \OO \to A^*$ which is extended to  \hom
$\sig: (A \cup \OO)^* \to A^*$ such that $\sig(U) = \sig(V)$.
Word equations are studied in other algebraic structures and frequently one is not interested only in satisfiability. 
For example, one may be  interested in all solutions, or only in solutions satisfying additional criteria 
like \emph{rational constraints} for free groups \cite{dgh05IC}. Here, we focus on the simplest case of word equations over free monoids; and by \emph{WordEquations} we understand the formal language of all word equations (over a given finite alphabet $A$) which are satisfiable, 
 that is, for which there exists a solution. 

\subsection*{History}

The problem WordEquations is closely related to the theory 
of Diophantine equations. The publication of Hilbert's 1900 address to the International Congress of Mathematicians listed 23 problems. The tenth problem (Hilbert 10) is:
\begin{quote}``Given a Diophantine equation with any number of unknown quantities and with rational integral numerical coefficients: To devise a process according to which it can be determined in a finite number of operations whether the equation is solvable in rational integers.''\end{quote}

There is a natural encoding of a word equation as a \Dio problem. 
It is based on the fact 
that two $2\times 2$ integer matrices 
$\vdmatrix1011$ and $\vdmatrix1101$ generate a free monoid.
Moreover, these matrices generate exactly those matrices on $\SL(2,\Z)$ 
where all coefficients are natural numbers. 
This is actually easy to show, and also used in fast ``fingerprint'' pattern matching algorithm by Karp and Rabin \cite{KarpRabin87}.
A reduction from WordEquations to Hilbert 10 is now straightforward.
For example, the equation $abX=Yba$ is solvable \IFF 
 the following \Dio system in unknowns ${X_1}$, \ldots,  ${Y_4}$ is solvable
 over integers:
\begin{align*}
\vdmatrix1011 \cdot \vdmatrix1101 \cdot \vdmatrix{X_1}{X_2}{X_3}{X_4}
&= 
 \vdmatrix{Y_1}{Y_2}{Y_3}{Y_4}\cdot \vdmatrix1101  \cdot \vdmatrix1011 \\
X_1 X_4 - X_2 X_3 &= 1\\
Y_1 Y_4 - Y_2 Y_3 &= 1\\
X_i\geq 0 \quad\& \quad Y_i\geq 0 & \quad \text{ for } 1 \leq i \leq 4\\
\end{align*}
The reduction of a \Dio system  to a single \Dio equation is classic. 
It is based on the fact that every natural number can be written as a sum of four squares. 
In the mid 1960s the following mathematical project was launched: 
show that Hilbert 10 is undecidable by showing that WordEquations is undecidable. 
The hope was to encode the computations of a Turing machine into a word equation. The project failed greatly, producing two great mathematical 
achievements. 
In 1970 Matiyasevich showed that Hilbert 10 is undecidable, based on
number theory and  previous work by Davis, Putnam, and Robinson, see the textbook \cite{mat93}. A few years later, in 1977 
Makanin showed that  WordEquations is decidable \cite{mak77}. 

In the 1980s, Makanin showed that the existential and positive theories of free groups  are decidable \cite{mak84}. In 1987 Razborov gave a description of all solutions for an equation in a free group via ``Makanin-Razborov'' diagrams \cite{raz87,raz93}. Finally, in a series of papers ending in  
\cite{KMIV06} Kharlampovich and Myasnikov proved Tarski's conjectures dating back to the 1940s:
\begin{enumerate}\item The elementary theory of free groups is decidable. 
\item Free non-abelian groups are elementary equivalent. 
\end{enumerate}
The second result has also been shown independently by Sela \cite{sela13}.

It is not difficult to see (by encoding linear \Dio systems over the naturals)  that WordEquations is NP-hard, but the first estimations of Makanin's algorithm was something like 
$$\text{DTIME}\left(2^{2^{2^{2^{2^{\mathrm{poly}(n)}}}}}\right).$$ 
Over the years Makanin's algorithm was modified to bring the complexity down to \EXPSPACE\ \cite{gut98focs}, see also the survey in \cite{die98lothaire}. 
For equations in free groups the complexity seemed to be much worse. Ko{\'s}cielski and Pacholski published a result  that the scheme of Makanin's algorithm  for free groups is not primitive recursive  \cite{kp96}.
However, a few years later  Plandowski and Rytter showed in \cite{pr98icalp} that 
solutions of word equations can be compressed by 
{L}empel-{Z}iv encodings (actually by straight-line programs); and 
the conjecture was born that WordEquations is in NP; and, moreover, the same should be true for word equations over free groups. 
The conjecture has not yet been proved, but in 1999 Plandowski showed that 
WordEquations is in \PSPACE\ \cite{pla99focs,pla04jacm}. The same is true 
for equations in free groups and allowing rational constraints we obtain a 
 \PSPACE-complete problem \cite{gut2000stoc,dgh05IC}.
 
 In 2013 Je\.z applied \emph{recompression} to WordEquations
and simplified all (!) known proofs for decidability \cite{jez13stacs}.
Actually, using his method he could describe all solutions of a word equation
by a finite graph where the labels are 
of two types. 
Either the label is a compression $c\mapsto ab$ where $a,b,c$ where letters
or the label is a linear \Dio system. His method 
 copes with free groups and with rational constraints: this was done 
 in \cite{DiekertJP2014csr}.

 Moreover, the method of Je\.z led Ciobanu, Elder, and the present author to an even simpler description for the set of all solutions: it is an EDT0L language \cite{CiobanuDEicalp2015}. Such a simple structural description of solution sets was known before only for quadratic word equations by \cite{FerteMarinSenizerguesTocs14}. 

 The notion of an {\em EDT0L system} refers to {\em {\bf E}xtended, {\bf D}eterministic, {\bf T}able, 
{\bf 0} interaction, and {\bf L}indenmayer}. 
There is a vast literature on Lindenmayer systems, see \cite{RozS86}, but actually we need very little from the ``Book of {\bf L}''.

\subsection*{Rational sets of \Endos}
The starting point is a word equation $(U,V)$ of length $n$ over a set of constants $A$  and set of variables $X_1, \ldots, X_k$ (without restriction,  $\abs A + k\leq n$). There is an nondeterministic algorithm which takes $(U,V)$ as input and which works in space $\NSPACE(n \log n)$. The output is   
an extended alphabet $C\supseteq A$ of linear size in $n$  and a finite trim 
nondeterministic automaton $\cA$ where the arc labels are \Endos over $C^*$.
The automaton $\cA$
accepts therefore a rational set $\cR = L(\cA) \sse \End(C^*)$, and 
 enjoys various properties which are explained next. 
The arc labels are restricted.
An \Endo used for an arc label is defined by mapping 
$c\mapsto u$ where $c\in C$ is a letter and 
$u$ is some word of length at most $2$.
The monoid  $\End(C^*)$ is neither free nor finitely generated, but 
$\cR$ lives inside a finitely generated submonoid $H^* \sse \End(C^*)$ where $H$ is finite. Thus, we can think of $\cR$ as a rational (or regular) expression over a finite set of \Endos $H$ as we are used to in standard formal language theory. For technical reasons it is convenient to assume that
$C$ contains a special symbol $\#$ whose main purpose is serve as a marker. 
The algorithm is designed in such a way that it yields an automaton $\cA$ 
accepting a rational set $\cR$ such that 
$$\set{h(\#)}{h \in \cR} \sse \underbrace{A^*\# \cdots  \# A^*}_{k-1 \text{ symbols } \#}.$$
Thus, applying the set of \Endos to the special symbol $\#$ we obtain a
formal language in $(A^* \os \#)^{k-1}A^*$. The set 
$\set{h(\#)}{h \in \cR}$ encodes a set of $k$-tuples over $A^*$. 
Due to Asfeld  \cite{Asveld1977} we can take a description like $\set{h(\#)}{h \in \cR}$
as the very  \emph{definition} for EDT0L.
Now, the result  by Ciobanu et al.~in \cite{CiobanuDEicalp2015} is the following equality: 
$$\set{h(\#)}{h \in \cR} = \set{\sig(X_1)\# \cdots \#\sig(X_k)}{\sig(U)=\sig(V)}.$$
Here, $\sig$ runs over all solutions of the equation $(U,V)$. Hence, 
the set of all solutions for a given word equation is an EDT0L language.

The results stated in \cite{CiobanuDEicalp2015} are more general.\footnote{Full proofs are in \cite{CiobanuDEarxiv2015}.} They cope with the existential theory of equations with rational constraints in
finitely generated free products of free groups, finite groups, free monoids, and free monoids with involution. For example, they cover the 
existential theory of equations with rational constraints in the modular group 
$\PSL(2,\Z)$. 

The $\NSPACE(n \log n)$ algorithm produces some $\cA$   
whether or not  
$(U,V)$ has a  solution. (If there is no \solu then the trimmed automaton $\cA$ has no states accepting the empty set.)
This shifts the viewpoint on how to solve equations. 
The idea is that $\cA$ answers basic questions  about the solution set of $(U,V)$. Indeed, 
the construction in \cite{CiobanuDEicalp2015} is such that the following assertions hold. 
\begin{itemize}
\item The equation $(U,V)$ is solvable \IFF $L(\cA) \neq \es$.
\item The equation $(U,V)$ has infinitely many solutions  \IFF $L(\cA)$ is infinite.
\end{itemize}
In particular, decision problems like ``Is $(U,V)$ satisfiable?''
 or ``Does $(U,V)$ have infinitely many solutions'' can be answered in 
$\NSPACE(n \log n)$ for finitely generated free products over free groups, finite groups, free monoids, and free monoids with involution. 
Actually, we conjecture that $\NSPACE(n \log n)$ is the best complexity bound for WordEquations with respect to space. This conjecture might hold even if the problem WordEquations was in \NP. 
\subsection*{How to solve a linear \Dio system}
Many of the aspects of  
our method of 
solving word equations are present in the special case of solving a system of word equations over a unary alphabet. 
In this particular case Je\.z's recompression is closely related to  \cite{BoudetC96}. There are many other places where the following is explained, so in some sense we can view the rest of this section as folklore. 

Assume that Alice wants to explain to somebody, say Bob, in a very short time, say 15 minutes, that the set of solvable linear \Dio systems over integers is decidable. Assume that this fundamental insight is entirely new to Bob. Alice might start to explain something with Cramer's rule, determinants or Gaussian elimination, 
but Bob does not know  any of these terms, so better not to start with 
a course on linear algebra within a time slot of 15 minutes. 
\goodbreak

What Bob knows are basic matrix 
operations
and the notion of 
a linear \Dio system:  $$ AX = c, \; \text{ where } A \in \Z^{n\times n},\;  X= (X_1,\ldots, X_n)^T\; \text{and} \; c\in \Z^{n\times 1}.$$
Here, the  $X_i$ are variables over natural numbers. (This is not essential, and actually 
makes the problem  
more difficult than looking for a solution over integers.) 
 
The complexity of the problem depends on the 
or values
$ n$, $ \Absone c = \sum_{i}{\abs {c_{i}}}$ and $\Absone A = \sum_{i,j}{\abs {a_{ij}}}.$
Without restriction (by adding dummies) we have 
\begin{equation}\label{eq:dio}
\Absone c \leq \Absone A .
\end{equation}

Alice explains the compression algorithm with respect to a given solution $x\in \N^{n}$. 
Of course, the algorithm does not know the solution, 
so  the  algorithm uses  nondeterministic guesses.
This is allowed provided  two properties are satisfied: soundness and completeness. Soundness means that a guess can never transform a 
unsolvable system into a solvable one. Completeness means that for 
every solution $x$, there is some choice of correct guesses such that the procedure terminates with a system which has a trivial solution. 

So we begin by guessing a solution  $x\in \N^{n}$. First, we can check whether $x= 0$ is a \solu by looking at $c$. Indeed,
$x= 0$ is a \solu \IFF $c=0$. 

Hence, let us assume 
$x \neq 0$ (this might be possible even if $c=0$.) We define a vector $b= c$. The vector  $b$ (and the solution $x$) will be modified during the procedure. 
Perform the following while-loop.\\

\noindent 
{\bf while} $x \neq 0$ {\do}
\begin{enumerate}
\item For all $i$ define $x_i' = x_i-1$ if $x_i$ is odd and $x_i' = x_i$ otherwise. Thus, all  $x_i'$ are even. Rewrite the system with a new vector
$b'$ such that 
$Ax'= b'.$
Note that 
\begin{equation}\label{eq:dioo}
\Absone{b'} \leq \Absone{b} + \Absone{A}.
\end{equation} 
\item Now, all $b_i'$ must be even. 
Otherwise we made a mistake and $x$ was not a solution.  
\item Define $b_i''= b_i'/2$ and $x_i''= x_i'/2$. 
We obtain a new system $AX= b''$ with solution $Ax''= b''.$
\item Rename $b''$ and $x''$ as $b$ and $x$.
\end{enumerate}
\noindent 
{\bf end while.}\\

 The clue
 is  that, since  $\Absone{b} \leq \Absone{A}$ by \prref{eq:dio},
 we obtain by \prref{eq:dioo} and the third step an invariant:
$$\Absone{b''}= \Absone{b'}/2 \leq \Absone{b}/2 + \Absone{A}/2 \leq \Absone{A}.$$
The procedure is obviously sound. 
It is complete because in each round $\Absone x$ decreases and therefore termination is guaranteed for every 
solution as long as we make correct guesses. The final observation is that the procedure defines a finite graph.
The vertices are  the vectors $b \in \Z^n$ with  $\Absone{b} \leq \Absone{A}.$
There are at most $\Absone{A}^{2n+1}$ such vectors. We are done! It is reported that the explanation of Alice took less than 15 minutes. It is not reported whether Bob understood. 

Alice explanation has a bonus: there is more information. We can label the arcs according to our guesses with affine mappings of two types: either $x \mapsto x+1_I$ or $x \mapsto 2x$. Here $1_I$ denotes the characteristic vector over a non-empty set $I \sse \os{1, \ldots, n}$. 

Thus, we have a finite graph of at most exponential size where the arc labels are affine mappings of type
$x \mapsto \lam x + 1_I$ with $\lam \in \os{1,2}$  and $I \sse \os{1, \ldots, n}$.
Letting $b=0$ be the initial state and the initial vector $c$ the final state,  we have  a nondeterministic finite automaton which accepts a rational set $\cR$ of affine mappings from $\N^n$ to itself. By construction, we obtain
$$\set{x \in \N^n}{Ax = c} = \set{h(0)}{h \in \cR}.$$

\newcommand{\Ju}{Ju}\newcommand{\Ph}{Ph}\newcommand{\Th}{Th}\newcommand{\Ch}{Ch}\newcommand{\Yu}{Yu}\newcommand{\Zh}{Zh}\newcommand{\St}{St}\newcommand{\curlybraces}[1]{\{#1\}}

\end{document}